\documentclass[twocolumn]{jpsj2}

\def\runauthor{J. \textsc{Otsuki} and Y. \textsc{Kuramoto}}

\title{%
Extension of the Non-Crossing Approximation for the Anderson Model\\ with Finite Coulomb Repulsion
}

\author{%
Junya \textsc{Otsuki}\thanks{E-mail address: otsuki@cmpt.phys.tohoku.ac.jp} 
and Yoshio \textsc{Kuramoto}
}

\inst{%
Department of Physics, Tohoku University, Sendai 980-8578
}

\recdate{\today}

\abst{%
The non-crossing approximation (NCA) is generalized 
for the Anderson model with finite Coulomb repulsion $U$. 
Resummation 
of infinite number of terms 
is performed so as to incorporate all leading contributions in the limit of large degeneracy of the local states. 
With negligible weight of the doubly occupied local states in equilibrium,
the extension is achieved through simple modifications 
of the lowest order formulae. 
The single-particle excitation spectrum is 
calculated with almost the same computational effort as in the 
original NCA.
The 
present scheme reproduces the scaling behavior 
of the Kondo resonance in the density of states, and 
gives 
reasonable description of thermodynamics
of the finite-$U$ Anderson model over a wide range of temperature. 
}

\kword{%
non-crossing approximation (NCA), Anderson model, single-particle excitation spectrum, entropy, specific heat, resolvent method
}

\begin{document}
\maketitle

\section{Introduction}
Heavy fermion systems, where $4f$ electrons 
play an important role, show plenty of fascinating phenomena. 
Such systems are 
often modeled by the periodic Anderson model, which exhibits the Kondo effect as well as the mixed-valence property.
The single-impurity Anderson model is 
utilized in the intermediate stage for solving the 
periodic model by means of the dynamical mean field theory (DMFT)\cite{Georges}.
Combined with band calculations, the DMFT can give a semi-quantitative description of density of states in real compounds\cite{LDA_DMFT1, LDA_DMFT2, Sakai2}.
For this purpose, it is required to solve the impurity model under realistic conditions for system parameters, which may lead to a heavy load in numerical calculation. 
Hence it is still important to develop a simple but reliable computational scheme to solve the impurity Anderson model, 
although 
low-energy properties of the model have now been understood well.

Perturbation theory with respect to the hybridization 
can take into account strong correlations between local electrons 
\cite{nca1, Grewe, Keiter-Morandi, Bickers, Coleman}. 
The lowest order self-consistent approximation is called the non-crossing approximation (NCA), since the scheme incorporates all diagrams without crossing of conduction lines.
In the case of large degeneracy $n$ of the local states, the NCA derives proper dynamics as well as thermodynamics of the Anderson model with infinite Coulomb repulsion $U$. 
The NCA accordingly plays a role 
complementary to the numerical renormalization group (NRG)\cite{Wilson} and the quantum Monte Carlo (QMC)\cite{Hirsch-Fye}, since 
the latter numerical methods are only available in the small $n$ case by restriction of computational time.
Combination of the NCA and the DMFT has so far revealed various characteristics of
the periodic Anderson model\cite{xnca1, xnca2} and the Hubbard model\cite{Pruschke_Hubbard}.

The NCA is originally derived for the infinite-$U$ Anderson model, where only $4f^0$ and $4f^1$ states are contained.
For practical use, however, additional informations on $4f^2$ states are necessary as well.
Extension of the NCA to the finite-$U$ case has been tried in several ways\cite{Sakai1, Pruschke, Saso, Kroha}.
It has been found that 
increase of the Kondo temperature, which should occur due to the involvement of $4f^2$ states, cannot be 
reproduced by lower order modification of the NCA to the finite-$U$ case, even though $n$ is large\cite{Sakai1}.
Therefore, higher order terms are relevant in such models.

Sakai {\it et al.} have pointed out importance of infinite number of skeleton diagrams in large-$n$ limit, and has 
called their approximation NCA$f^2$v\cite{Sakai1}.
The theory seems to yield a proper energy scale of the Kondo effect. 
However, the resummation 
procedure for the single-particle excitation spectrum 
does not give a conserving approximation.
Haule {\it et al.} have developed a conserving approximation by selecting a set of generating functionals\cite{Kroha}.
It is called symmetrized finite-$U$ NCA (SUNCA), since the approximation incorporates excitations of 
both particles and holes on equal footing,
and can accordingly reproduce the symmetric feature of the single-particle excitation spectrum. 
Although the theory provides a proper energy scale in $n=2$, 
actual computation is not easy for large $n$ even for a simple structure of the conduction band. 

In order to be applied to realistic impurity models with 
complex $4f$-shell and band structures, or to the DMFT as a solver of the single-impurity problem, 
the practical calculational scheme should be simple enough.
In this paper, we 
modify the NCA$f^2$v scheme so as to recover the conserving property in the limit of large $n$,
and present 
results for the single-particle spectra. 
The computations 
require almost the same effort as in the original NCA.
It will be shown that the theory gives a proper energy scale and characteristics of the Kondo effect under the large-$n$ condition. 
In addition we check the accuracy of the theory in thermodynamics.

This paper is organized as follows.
Section 2 introduces a model that we will consider in this paper, and gives a brief description of the resolvent method.
Formulations for the single-particle excitation are given in \S 3. 
In \S 4, we present numerical results, and discuss their accuracy in several aspects.
In the final section, we summarize characteristics of the present theory.

\section{Model and Resolvent Method}
We consider the Anderson model with 
an orbital degeneracy in the local states, which are called $4f$ states hereafter.
Assuming strong Hund's coupling,
we distinguish the $4f$ states by a single quantum number $\alpha$ with $n$-fold degeneracy, e.g., the azimuthal component of the total angular momentum $j=5/2$. 
Conduction electrons 
hybridize with $4f$ electrons having the same index $\alpha$.
Then, the Hamiltonian takes the form
\begin{align}
	H &= \sum_{\mib{k} \alpha} \epsilon_{\mib{k}} c^{\dag}_{\mib{k} \alpha} c_{\mib{k} \alpha}
	 + \epsilon_f \sum_{\alpha} f^{\dag}_{\alpha} f_{\alpha} \nonumber \\
	 & \quad + \frac{U}{2} \sum_{\alpha \neq \alpha'} f^{\dag}_{\alpha} f_{\alpha} f^{\dag}_{\alpha'} f_{\alpha'}
	 + \sum_{\mib{k} \alpha} V_{\mib{k}} f^{\dag}_{\alpha} c_{\mib{k} \alpha} + \text{h.c.},
\end{align}
where $c_{\mib{k} \alpha}$ and $f_{\alpha}$ is annihilation operators for the conduction electron and the $4f$ electron, respectively.

We 
work under the large-$U$ condition, and accordingly restrict $4f$ states to $4f^0$, $4f^1$ and $4f^2$. 
In order to take account of the strong correlation between $4f$ electrons, we introduce the resolvent $R_{\gamma}(z)$ for each $4f$ state as follows:
\begin{align}
	R_{\gamma} (z) = [z -\epsilon_{\gamma} - \Sigma_{\gamma}(z)]^{-1},
	\label{eq:resolv}
\end{align}
where $\gamma$ 
indicates either $4f^0$, $4f^1$ or $4f^2$ states, and hereafter we use corresponding indices 0, 1 or 2, respectively. $\epsilon_{\gamma}$ is the unperturbed energy, i.e., $\epsilon_0 =0$, $\epsilon_1 = \epsilon_f$ and $\epsilon_2 =2\epsilon_f + U$.
The hybridization 
is to be included in the self-energy part $\Sigma_{\gamma}(z)$ by the self-consistent perturbation theory\cite{nca1, Grewe, Keiter-Morandi, Bickers}.

Once the resolvent is obtained, physical quantities are computed with use of them\cite{nca1, Grewe, Keiter-Morandi, Bickers}.
For example, the partition function $Z_f$ of $4f$ part is given by 
\begin{align}
	Z_f = \int_C \frac{\text{d}z}{2\pi \text{i}} \ e^{-\beta z} \sum_{\gamma} R_{\gamma} (z),
	\label{eq:part_func}
\end{align}
where the contour $C$ encircles all the singularities of $R_{\gamma}(z)$ counter-clockwise. 
The resolvent is utilized to calculate dynamics as well.

Throughout this paper, we give formulations on 
the assumption that 
there is no splitting in 
$4f^1$ or $4f^2$ levels for simplicity.
Alternative formulations with such splittings are obtained 
straightforwardly by distinguishing resolvents for each split level.

%

\section{Finite-$U$ Perturbation Theory}
The excitations of the conduction holes 
always accompany a summation of $4f^1$ or $4f^2$ states, which produces a factor $n$ or $n-1$, respectively. 
Hence from viewpoint of the $1/n$ expansion, the hole excitations
are not in higher order in perturbation series,
and proper account should be taken of consecutive occurrences of such excitations.
Since $4f^0$-$4f^1$ fluctuation terms are represented by the diagrams without crossing of conduction lines, they are counted by the self-consistent treatment of the lowest order skeleton diagrams.
On the other hand, $4f^1$-$4f^2$ fluctuations involve crossing of conduction lines. 
Consequently, 
higher order skeleton diagrams are necessary to take account of such terms.

In this section, we present an extended theory where $4f^1$-$4f^2$ fluctuations 
accompanying conduction holes are included properly. 
We first review the theory by Sakai {\it et al.} where a vertex part is introduced for the self-energy.
After a brief description is given of a relation to the theory by Haule {\it et al.},
we proceed to the formulation of the single-particle excitation.

\subsection{The NCA$f^2$v scheme}
\begin{figure}[t]
	\begin{center}
	\includegraphics[width=8.5cm]{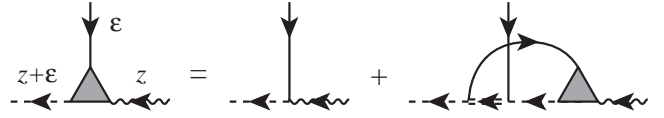}
	\end{center}
	\caption{Diagrammatical representation of the equation for the vertex function. Solid, wavy,  dashed and double dashed line represent conduction electron, $4f^0$, $4f^1$ and $4f^2$ resolvent, respectively. All resolvents are regarded as renormalized ones.}
	\label{fig:vertex_eq}
\end{figure}

Before we present formulations for the single-particle excitation, we summarize 
the NCA$f^2$v scheme developed by Sakai {\it et al.}\cite{Sakai1, Sakai2}
In order to derive the self-energy, they have introduced the 
vertex part $\Lambda(z, \epsilon)$
which consists of the $4f^1$-$4f^2$ fluctuations with consecutive excitations of conduction holes:
\begin{align}
	\Lambda(z, \epsilon) = 1 + (n-1) \int {\rm d}\epsilon' &W(\epsilon') f(\epsilon')
	 R_1(z+\epsilon') \nonumber \\
	 &\times R_2(z+\epsilon+\epsilon') \Lambda (z, \epsilon'),
	\label{eq:vertex_eq}
\end{align}
where $f(\epsilon)$ is the Fermi distribution function and $W(\epsilon)$ is defined by $W(\epsilon)=\sum_{\mib{k}} |V_{\mib{k}}|^2 \delta(\epsilon - \epsilon_{\mib{k}})$. 
Equation~(\ref{eq:vertex_eq}) 
is diagrammatically represented by Fig.~\ref{fig:vertex_eq}. The vertex part can 
equally be represented by diagrams with all arrows reversed in Fig.~\ref{fig:vertex_eq}.
%

A computation of the function $\Lambda(z, \epsilon)$ with double argument takes a lot of time.
Sakai {\it et al.} have pointed out that the energy dependence of $R_2 (z)$ is less important than $R_1 (z)$'s in eq.~(\ref{eq:vertex_eq}), and 
proposed an approximation where $R_2 (z)$ is replaced by a constant value $\tilde{R}_2 = R_2^{(0)}(\epsilon_f)$, which is the bare resolvent of $4f^2$ state at $z=\epsilon_f$. 
In this approximation, the vertex part is independent of the energy of conduction 
electrons, and becomes a function of single argument written as $\Lambda(z)$. 
Equation~(\ref{eq:vertex_eq}) is 
then solvable and leads to an expression
\begin{align}
	\Lambda(z) = [1 - (n-1) \tilde{R}_2 \tilde{\Sigma}^{\rm (NCA)}_0 (z)]^{-1},
	\label{eq:vertex}
\end{align}
where $\tilde{\Sigma}^{\rm (NCA)}_0 (z)$ is defined by
\begin{align}
	\tilde{\Sigma}^{\rm (NCA)}_0 (z) = \int {\rm d}\epsilon W(\epsilon) f(\epsilon) R_1(z+\epsilon),
	\label{eq:self_nca}
\end{align}
which is the self-energy of $4f^0$ state in the NCA divided by $n$.
The approximated vertex function $\Lambda (z)$ is now analytic 
in the region except for ${\rm Im}z=0$, while original one $\Lambda (z, \epsilon)$ has the additional cut at ${\rm Im}(z+\epsilon)=0$ 
coming from $R_2(z)$.

\begin{figure}[t]
	\begin{center}
	\includegraphics[width=8cm]{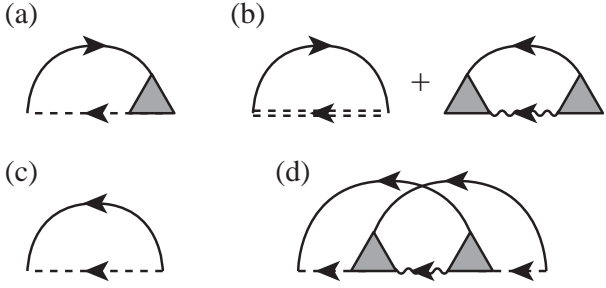}
	\end{center}
	\caption{Diagrammatical representation of the self-energy (a) $\Sigma_0 (z)$, (b) $\Sigma_1 (z)$, (c) $\Sigma_2 (z)$, and (d) a correction term which 
	should be included in $\Sigma_2 (z)$ for construction of a 
	conserving approximation.}
\label{fig:self-energy}
\end{figure}

The self-energy part of each state is represented with use of the vertex part introduced above. 
The self-energy of $4f^0$ state is obtained by multiplying the NCA self-energy and the vertex function $\Lambda(z)$ as follows:
\begin{align}
	\Sigma_0 (z) = n \tilde{\Sigma}^{\rm (NCA)}_0 (z) \Lambda(z).
	\label{eq:self0}
\end{align}
Corresponding diagrammatical representation is given by Fig.~\ref{fig:self-energy}(a).
Figure~\ref{fig:self-energy}(b) shows a diagram of $4f^1$ 
self-energy, whose analytic expression is
\begin{align}
	\Sigma_1 (z) &= \int {\rm d}\epsilon W(\epsilon) [1-f(\epsilon)] \tilde{R}_0(z-\epsilon) \nonumber \\
	& \quad + (n-1) \int {\rm d}\epsilon W(\epsilon) f(\epsilon) R_2(z+\epsilon),
	\label{eq:self1}
\end{align}
where $\tilde{R}_0(z)$ is defined by
\begin{align}
	\tilde{R}_0(z) = R_0(z) \Lambda(z)^2.
	\label{eq:R0_tilde}
\end{align}
Introduction of the auxiliary function $\tilde{R}_0(z)$ makes eq.~(\ref{eq:self1}) 
look similar to the lowest order self-energy. 
The self-energy of $4f^2$ states are evaluated without any vertex corrections by 
\begin{align}
	\Sigma_2 (z) = 2\int {\rm d}\epsilon W(\epsilon) [1-f(\epsilon)] R_1(z-\epsilon),
	\label{eq:self2}
\end{align}
which is expressed by the diagram of Fig.~\ref{fig:self-energy}(c). 
The term shown in Fig.~\ref{fig:self-energy}(d) is actually required for a construction of the conserving approximation as we will discuss later. 
We neglect the term, however, since its contribution is of order $1/n^2$ at most.
Thus we avoid the double integral with respect to the energy of conduction electrons that would take much time.
The self-energy in the NCA$f^2$v consequently contains all diagrams of the leading contributions in the large-$n$ limit, while the original NCA for the infinite-$U$ model incorporates diagrams up to the first order of $1/n$.

Resultant NCA$f^2$v equations have 
forms quite similar 
to that of the simplest finite-$U$ extension 
in spite of the resummation of all the leading contributions.
The only modification of the equations is the replacement of $\Sigma_0(z)$ 
and $R_0(z)$ by $\Sigma_0(z) \Lambda(z)$ and $R_0(z) \Lambda(z)^2$, respectively.
The vertex function $\Lambda(z)$ has been given analytically by eq.~(\ref{eq:vertex}). 
Accordingly, the computational effort is almost the same as in the lowest order self-consistent approximation without vertex corrections.

\subsection{Relation to the SUNCA}
Haule and coworkers have extended the NCA to the finite-$U$ 
case, and called the scheme SUNCA\cite{Kroha}. 
They have constructed a conserving scheme by taking a set of infinite number of generating functionals into consideration. 
We discuss relation between the SUNCA and the NCA$f^2$v. 

\begin{figure}[t]
	\begin{center}
	\includegraphics[width=5.5cm]{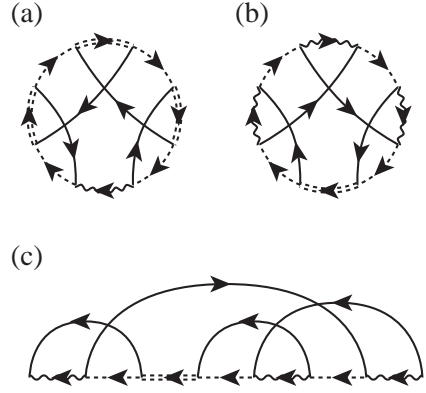}
	\end{center}
	\caption{Eighth order diagrams of the generating functionals (a) considered in both the SUNCA and the NCA$f^2$v, and (b) considered in the SUNCA but not in the NCA$f^2$v. Diagram (c) represents one of contributions to the $4f^1$ state self-energy derived from the functional (b), which is of order $1/n^3$.}
	\label{fig:functional}
\end{figure}

The generating functionals in the SUNCA consist of two parts except for terms less than the sixth order.
Figures~\ref{fig:functional}(a) and (b) show diagrams of the eighth order generating functionals that the SUNCA incorporates.
The functional (a) generates 
the self-energy terms included in the NCA$f^2$v. 
On the other hand, the NCA$f^2$v does not incorporate contributions generated from the functional (b).
Figure~\ref{fig:functional}(c) shows one of the eighth order terms of the $4f^1$ state self-energy derived from the functional (b). 
It includes three conduction electrons propagating forward, 
and accordingly is of order $1/n^3$. 
Corresponding vertex part 
coming from the (b)-type 
generating functionals consists of the $4f^0$-$4f^1$ fluctuations with consecutive excitations of forward propagating conduction electrons. 
Hence perturbation series of the self-energy generated from the (b)-type functionals are of order $1/n$ at most. 
Consequently, large-$n$ condition justifies the neglect of (b)-type diagrams, 
which are relevant in the small-$n$ case, especially $n=2$.

It is noticed that the NCA$f^2$v 
itself is a conserving scheme provided no further approximation is made such as
simplification of the vertex part and neglect of the correction term for $\Sigma_2 (z)$.
Since these approximations are justified 
under the condition of large-$n$ and large-$U$, 
we expect that violation of the conserving property is not serious under such condition.

\subsection{Single-particle Green function}
We now proceed to 
the single-particle excitation.
The $4f$ electron Green function is represented in terms of the
$X$-operator $X_{\gamma \gamma'} = |\gamma \rangle \langle \gamma'|$
as follows:
\begin{align}
	G_f (\tau) = - \sum_{\gamma \gamma' \gamma'' \gamma'''}
	 &\langle \gamma | f_{\alpha} |\gamma' \rangle
	 \langle \gamma'' | f^{\dag}_{\alpha} |\gamma''' \rangle \nonumber \\
	 &\times \langle T_{\tau} X_{\gamma \gamma'}(\tau) X_{\gamma'' \gamma'''} \rangle,
	\label{eq:def_green_func}
\end{align}
where $X_{\gamma \gamma'}(\tau)$ is 
in the Heisenberg picture. 
The Green function is now independent of quantum number $\alpha$.
After the Fourier transformation, the Green function $G_f({\rm i}\epsilon_n )$ is to be 
derived with use of the resolvents\cite{nca1, Grewe, Keiter-Morandi, Bickers}, where $\epsilon_n =(2n+1)\pi T$ is the Matsubara frequency.

\begin{figure}[t]
	\begin{center}
	\includegraphics[width=8.5cm]{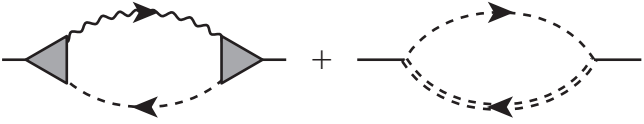}
	\end{center}
	\caption{Diagrammatical representation of the $4f$ electron Green function. }
	\label{fig:green_func}
\end{figure}

In order to construct a 
consistent approximation, infinite terms should be considered in the single-particle Green function as in the self-energy. 
The original NCA$f^2$v scheme, however, does not take proper account of the vertex correction. 
An appropriate formula of the Green function is given by
\begin{align}
	G_f &({\rm i}\epsilon_n) = \frac{1}{Z_f} \int_C \frac{dz}{2\pi \text{i}} e^{-\beta z}
	 \tilde{R}_0 (z) R_1 (z+{\rm i}\epsilon_n) \nonumber \\
	&+ (n-1) \frac{1}{Z_f} \int_C \frac{dz}{2\pi \text{i}} e^{-\beta z}
	 R_1 (z) R_2 (z+{\rm i}\epsilon_n),
	\label{eq:green_func}
\end{align}
where the contour $C$ encircles $\text{Im}z=0$ and $\text{Im}z=-{\rm i}\epsilon_n$ counter-clockwise, and $\tilde{R}_0 (z)$ is defined by eq.~(\ref{eq:R0_tilde}). 
Since the vertex function $\Lambda(z)$ is independent of the energy of the conduction electron and has the same analyticity as $R_0 (z)$ in the present approximation, the Green function has been given 
simply by 
replacing $R_0 (z)$ with $\tilde{R}_0 (z)$ in the lowest order formula. 
Corresponding diagram is represented by Fig.~\ref{fig:green_func}. 
We note that the first term of right-hand side of eq.~(\ref{eq:green_func}) 
contains not only diagonal components 
$\langle T_{\tau} X_{01}(\tau) X_{10} \rangle$ and $\langle T_{\tau} X_{12}(\tau) X_{21} \rangle$, 
but also off-diagonal components 
$\langle T_{\tau} X_{12}(\tau) X_{10} \rangle$ and $\langle T_{\tau} X_{01}(\tau) X_{21} \rangle$ 
when the vertex part is expanded.

Performing the integral with respect to $z$ and analytic continuation ${\rm i}\epsilon_n \rightarrow \omega +{\rm i}\delta$ onto real frequencies in eq.~(\ref{eq:green_func}), we obtain the single-particle 
excitation spectrum
$\rho_f(\omega)=-\pi^{-1} \text{Im}G_f (\omega+{\rm i}\delta)$ as follows:
\begin{align}
	&\rho_f (\omega)
	 = \int {\rm d}\epsilon[ \tilde{\xi}_0 (\epsilon) \eta_1(\epsilon+\omega)
	  +\tilde{\eta}_0 (\epsilon) \xi_1(\epsilon+\omega)] \nonumber \\
	 &+ (n-1) \int {\rm d}\epsilon[ \xi_1 (\epsilon) \eta_2(\epsilon+\omega)
	  +\eta_1 (\epsilon) \xi_2(\epsilon+\omega)]. 
	\label{eq:dos_4f}
\end{align}
We have included spectral functions of the resolvent $\eta_{\gamma}(\omega)$ and those of the defect propagator $\xi_{\gamma}(\omega)$ defined by 
\begin{align}
	\eta_{\gamma}(\omega) &= -\pi^{-1} \text{Im}R_{\gamma}(\omega+\text{i} \delta), 
	\label{eq:eta} \\
	\xi_{\gamma}(\omega) &= Z_f^{-1} \text{e}^{-\beta \omega} \eta_{\gamma} (\omega+ \text{i}\delta).
	\label{eq:xi}
\end{align}
Analogously, $\tilde{\eta}_0(\omega)$ and $\tilde{\xi}_0(\omega)$ are defined from $\tilde{R}_0(z)$.
Since it is 
awkward to compute $\xi_{\gamma}(\omega)$ by eq.~(\ref{eq:xi}) due to the Boltzmann factor, $\xi_{\gamma}(\omega)$ should be calculated by another set of equations 
which are obtained by transforming the original ones (see Appendix).

\subsection{Sum-rules}
The single-particle spectrum obtained within conserving scheme satisfies the following exact sum-rules:
\begin{align}
	&\int {\rm d}\omega f(\omega) \rho_f (\omega) = \langle X_{11} \rangle + (n-1) \langle X_{22} \rangle = \frac{n_f}{n}, \label{eq:4f_number} \\
	&\int {\rm d}\omega [1-f(\omega)] \rho_f (\omega) = \langle X_{00} \rangle + (n-1) \langle X_{11} \rangle,
	\label{eq:4f_number2}
\end{align}
where $n_f$ is the 
average number of $4f$ electrons. 
$\langle X_{\gamma \gamma} \rangle$ is an average occupation 
of $\gamma$ state, and is given in terms of $\xi_{\gamma}(\omega)$ by
\begin{align}
	\langle X_{\gamma \gamma} \rangle = \int \text{d}\omega \ \xi_{\gamma} (\omega).
	\label{eq:X_ave}
\end{align}
The sum-rule of eq.~(\ref{eq:4f_number}) follows from the fact that the left-hand side
corresponds to $G_f (-0)$ in eq.~(\ref{eq:def_green_func}).
Similarly the left-hand side of eq.~(\ref{eq:4f_number2}) corresponds to $G_f (+0)$.
Equation~(\ref{eq:4f_number}) has an intuitive interpretation that the $4f^1$ states are occupied in proportion to $n \langle X_{11} \rangle$, and the $4f^2$ states in proportion to $n(n-1) \langle X_{22} \rangle /2$. 
Equations~(\ref{eq:4f_number}) and (\ref{eq:4f_number2}) lead to 
another sum-rule of the form
\begin{align}
	\int {\rm d}\omega \rho_f (\omega) = \langle X_{00} \rangle + n\langle X_{11} \rangle
	 + (n-1)\langle X_{22} \rangle.
	\label{eq:dos_sum}
\end{align}
The sum-rule 
for $\xi_{\gamma} (\omega)$, given by eq.~(\ref{eq:sum_xi}), ensures that the 
integral of $\rho_f(\omega)$ 
is always less than unity. 
It is clear that the lowest order formula of the Green function satisfies the sum-rule, eqs.~(\ref{eq:4f_number}), (\ref{eq:4f_number2}) and (\ref{eq:dos_sum}), as in the original NCA\cite{nca1}. 
The fact may imply that any correction terms in the Green function should 
not affect the integrated spectral intensity.

We now discuss whether the density of 
states given by eq.~(\ref{eq:dos_4f}) 
satisfies the exact sum-rule in the present theory. 
We can see from eqs.~(\ref{eq:vertex}) and (\ref{eq:R0_tilde}) that $\tilde{R}_0(z)$ behaves at infinity as $\lim_{|z| \rightarrow \infty}\tilde{R}_0 (z)=z^{-1}$, similarly to $R_{\gamma}(z)$.
The asymptotic behavior leads to the following sum-rule for $\tilde{R}_0(z)$:
\begin{align}
	\int {\rm d}\omega \ \tilde{\eta}_0 (\omega) =1.
	\label{eq:sum_eta_tilde}
\end{align}
Equation~(\ref{eq:sum_eta_tilde}) ensures that the particle number is not affected by the vertex correction, and is given by eq.~(\ref{eq:4f_number}).
On the other hand, the approximate vertex correction changes the total intensity. 
Equation~(\ref{eq:4f_number2}) is 
modified due to the vertex correction by
\begin{align}
	\int {\rm d}\omega [1-f(\omega)] \rho_f (\omega) = \langle \tilde{X}_{00} \rangle + (n-1) \langle X_{11} \rangle,
	\label{eq:f2v_4f_number2}
\end{align}
where $\langle \tilde{X}_{00} \rangle$ is defined with use of $\tilde{\xi}_0 (\epsilon)$ in 
a manner similar to eq.~(\ref{eq:X_ave}).
Equation~(\ref{eq:f2v_4f_number2}) follows from eq.~(\ref{eq:dos_4f}).
If $\langle \tilde{X}_{00} \rangle$ were the same as 
$\langle X_{00} \rangle$, the exact sum-rule in eqs.~(\ref{eq:4f_number}), (\ref{eq:4f_number2}) and (\ref{eq:dos_sum}) would be satisfied.
The spectral intensity in our framework is examined by numerical calculation in the next section.

\section{Numerical Results and Discussions}
In this section, we present numerical results for single-particle excitation spectrum and thermodynamic quantities. 
Then we shall examine accuracy of the present theory.
We take a rectangular model $W(\epsilon) = W_0 \ \theta(D-|\epsilon|)$ with half width of conduction band being $D$.
The degeneracy $n$ takes $n=6$ and $n=4$. 
We take a unit such that $D=10^4$ throughout this paper. 
One may imagine the energy unit as of the order of Kelvin to meV for correspondence with actual systems.
Parameters $W_0$, $\epsilon_f$ and $U$ are varied in the range where the $4f$ particle number is almost unity. 

\subsection{Single-particle excitation spectrum}
We first show numerical results for density of states $\rho_f(\omega)$ of $4f$ electrons.
In addition to results in the present scheme, results in the other schemes are 
shown for comparison: 
(i) $\rho_f^{\rm (S)}$, 
which considers the vertex correction in the self-energy 
of resolvents but not in the Green function\cite{Sakai2,Sakai1},
(ii) NCA$f^2$, which is the lowest order self-consistent approximation including 
$4f^2$ states but without the vertex correction, and 
(iii) infinite-$U$ NCA, 
which is the original NCA with infinite Coulomb repulsion.

\begin{figure}[t]
	\begin{center}
	\includegraphics[width=8.5cm]{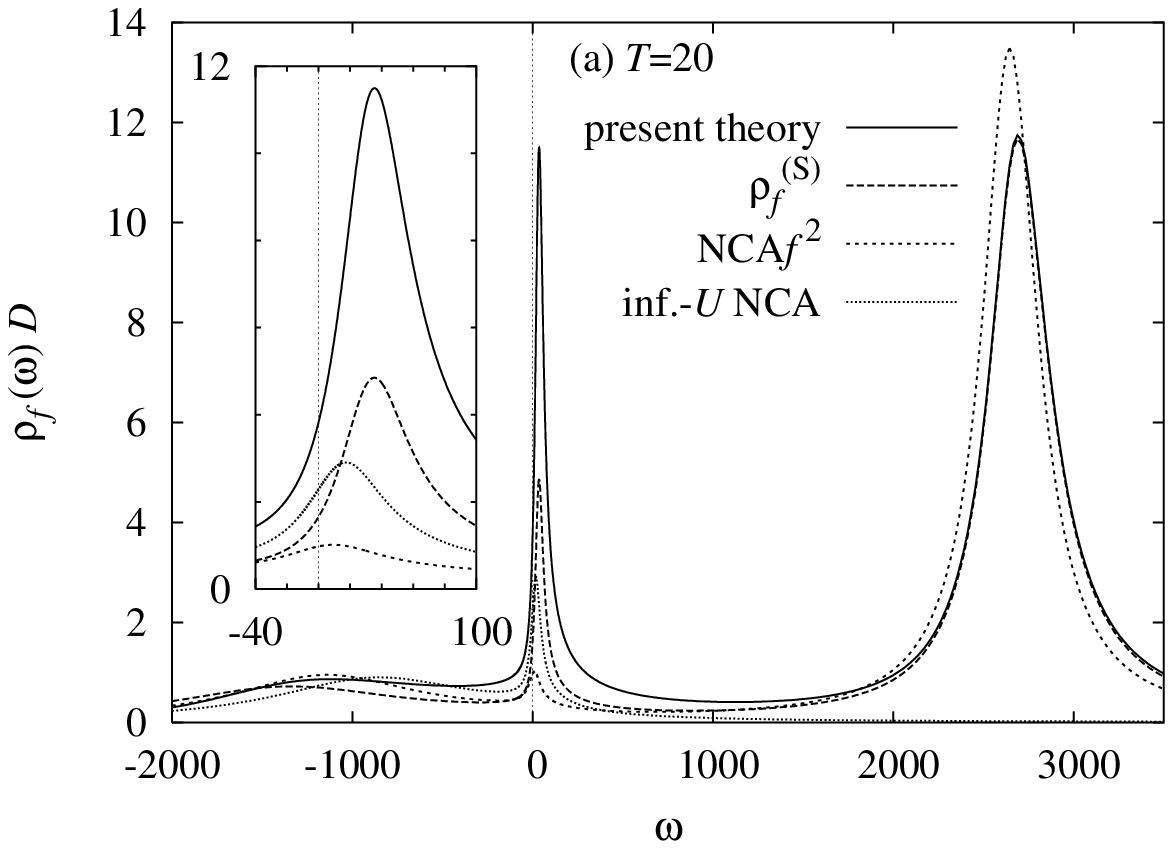}
	\includegraphics[width=8.5cm]{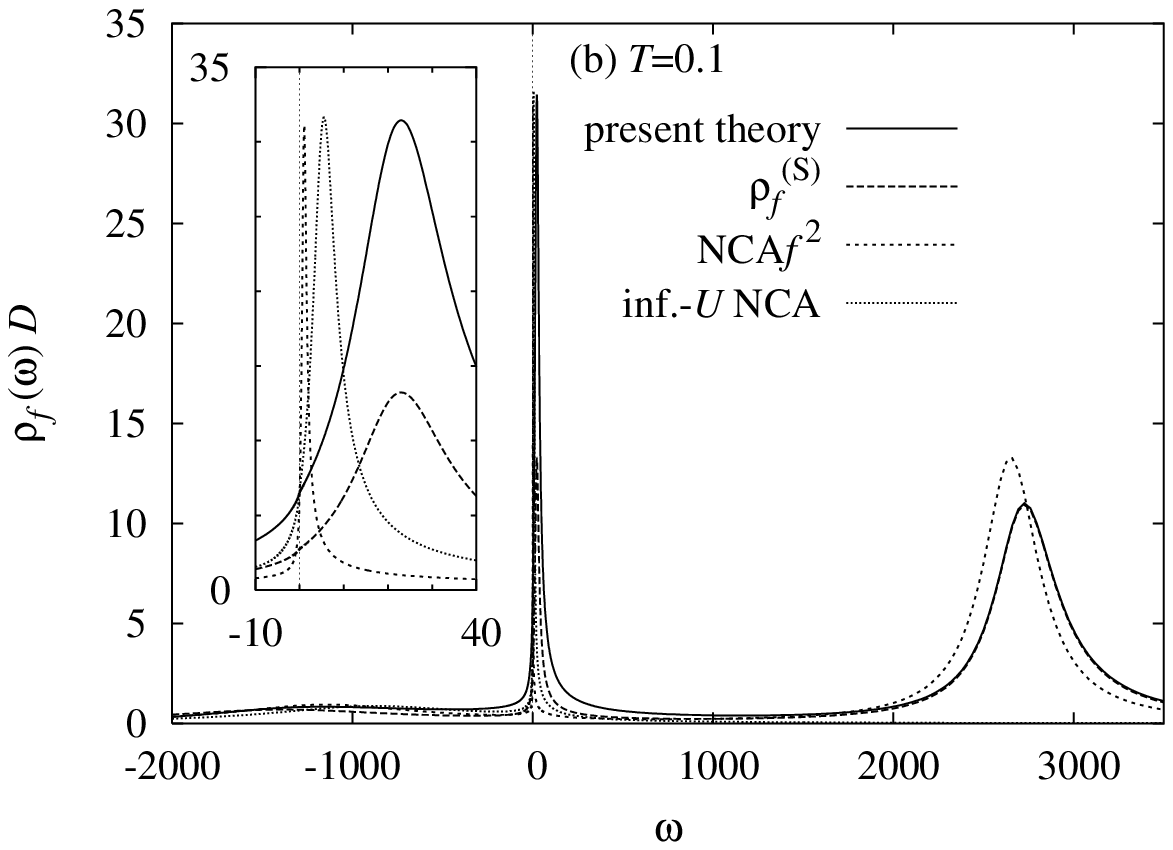}
	\end{center}
	\caption{Single-particle spectra computed in the several schemes with parameters $n=6$, $W_0=30$, $\epsilon_f = -1200$ and $U=3|\epsilon_f|=3600$ at temperatures (a) $T=20$ and (b) $T=0.1$.}
	\label{fig:dos_U3600}
\end{figure}

Figure~\ref{fig:dos_U3600} shows single-particle excitation with parameters $n=6$, $W_0=30$, $\epsilon_f =-1200$ and $U=3 |\epsilon_f|=3600$ at temperatures $T=20$ and $T=0.1$. 
In the NCA$f^2$, 
the peak of $4f^2$ states is slightly sharper and is located at lower energy than the other schemes.
There is 
no significant difference between the present theory and $\rho_f^{\rm (S)}$ so far as large $\omega$ is concerned.
Near the Fermi level, on the other hand, the heights of the Kondo resonance are considerably different from each other as shown in the inset of Fig.~\ref{fig:dos_U3600}(a).
This is mainly due to different values for the Kondo temperature $T_{\rm K}$ in different approximation schemes.
At much lower temperature $T=0.1$,
we can actually see from Fig.~\ref{fig:dos_U3600}(b) that the heights of the Kondo resonance peak are almost the same, except for $\rho_f^{\rm (S)}$. 
It is known that the peak position with large $n$ roughly gives $T_{\rm K}$ \cite{Kuramoto-MH}, 
which also determines the zero temperature limit of the susceptibility and the Sommerfeld coefficient of the specific heat. 
The precise definition of $T_{\rm K}$  is discussed in the next subsection.
We notice that $T_{\rm K}$ in the present theory is larger than that in the infinite-$U$ NCA as we expect, while the energy scale becomes smaller in the NCA$f^2$. 
The inaccuracy of the NCA$f^2$ scheme has been pointed out in ref.~\citen{Sakai2}.

\begin{figure}[t]
	\begin{center}
	\includegraphics[width=8.5cm]{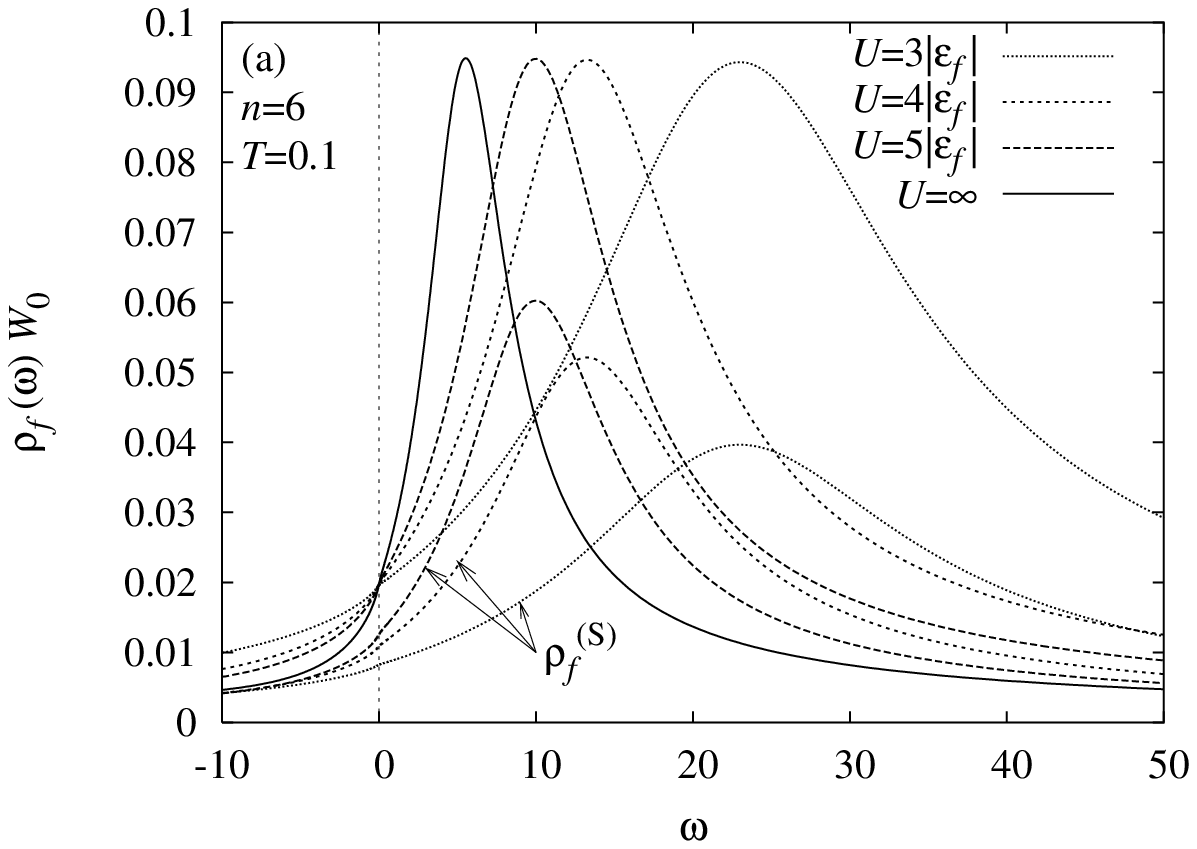}
	\includegraphics[width=8.5cm]{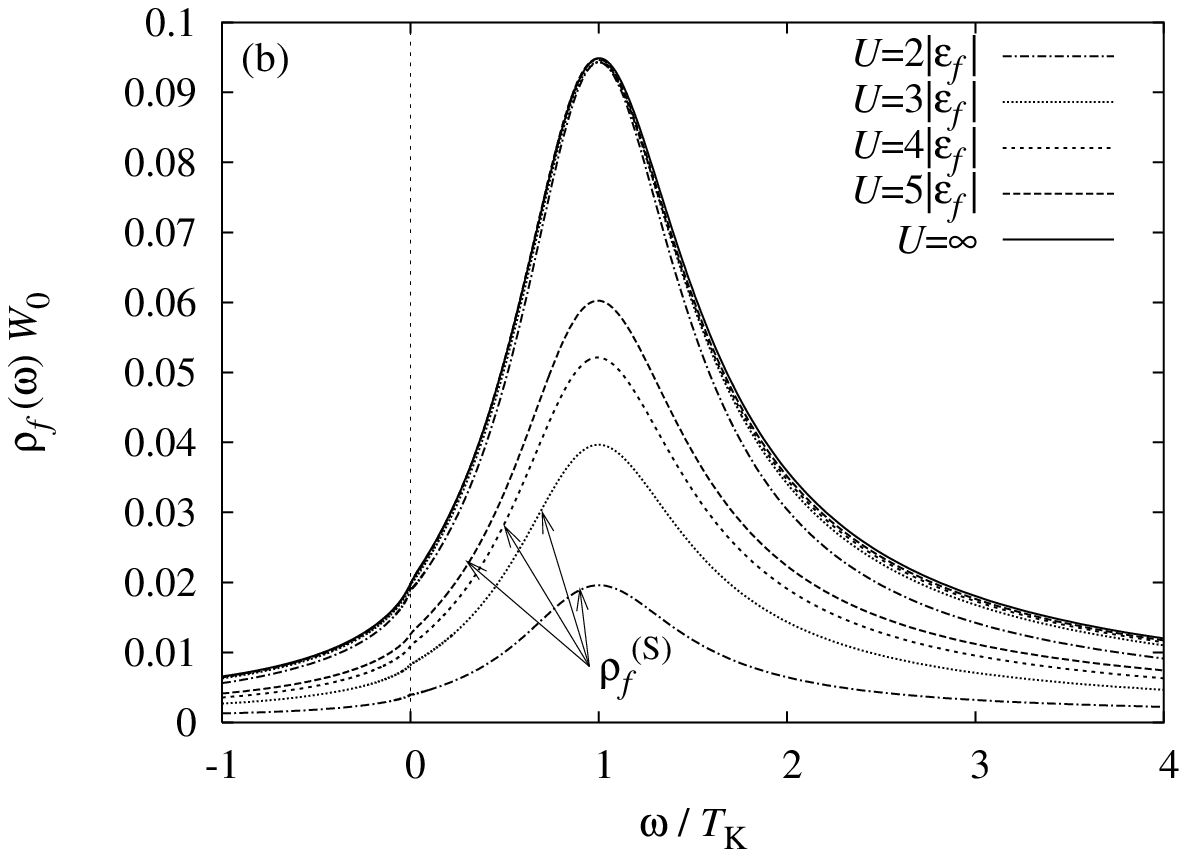}
	\end{center}
	\caption{Single-particle spectra near the Fermi level for several values of $U$ with $n=6$, $W_0 =30$ and $\epsilon_f = -1200$ at $T=0.1$. }
	\label{fig:dos_n6}
\end{figure}

In order to look through the Kondo behavior in this scheme, we show spectra near the Fermi level at low enough temperatures compared with the Kondo energy scale. 
Figure~\ref{fig:dos_n6}(a) shows spectra for several values of $U$ with $n=6$.
\begin{figure}[t]
	\begin{center}
	\includegraphics[width=8.5cm]{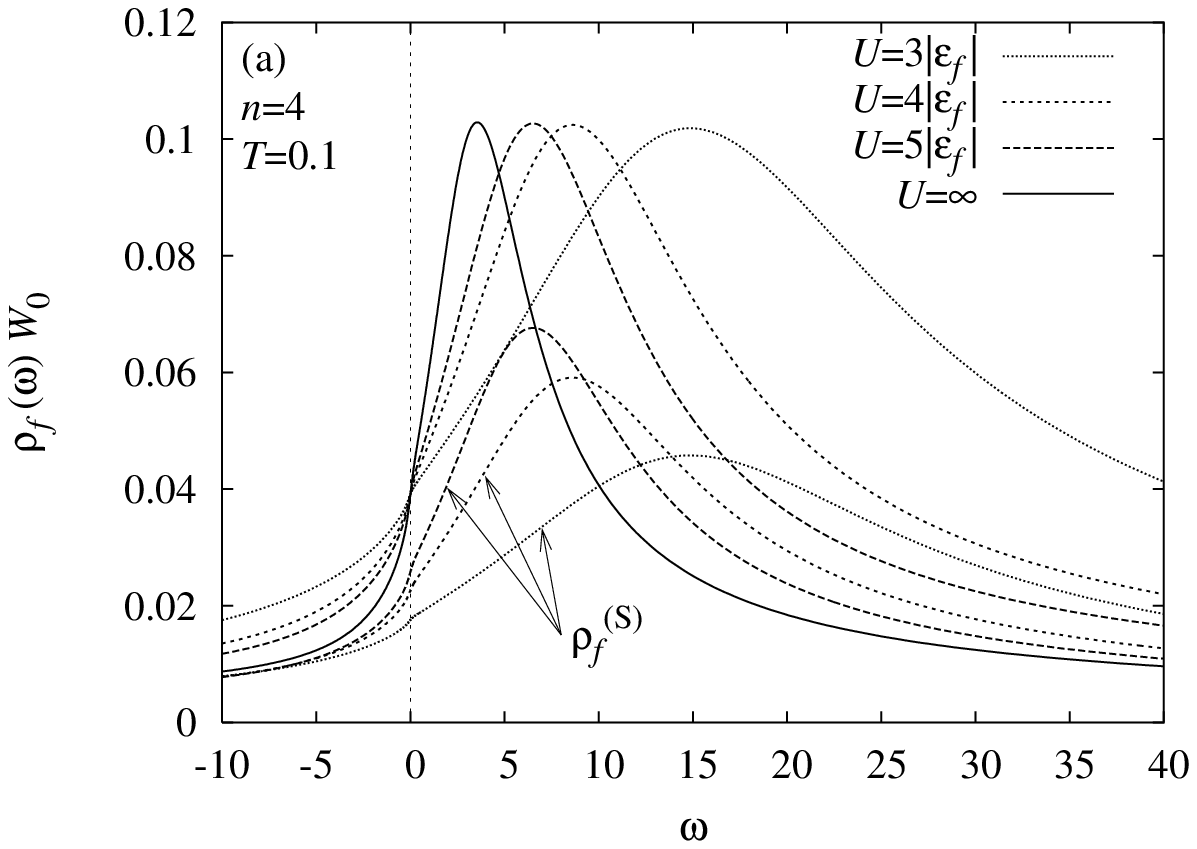}
	\includegraphics[width=8.5cm]{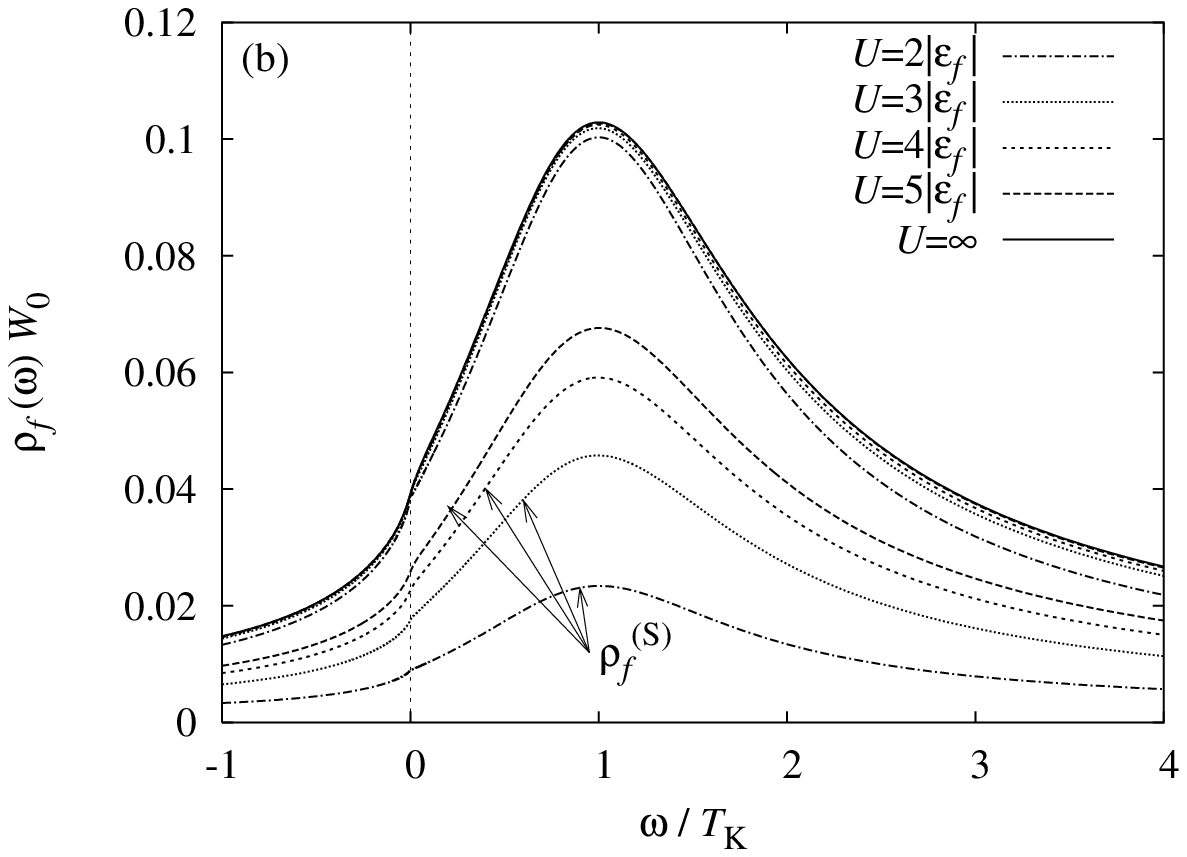}
	\end{center}
	\caption{Single-particle spectra near the Fermi level for several values of $U$ with $n=4$, $W_0 =45$ and $\epsilon_f = -1200$ at $T=0.1$. }
	\label{fig:dos_n4}
\end{figure}
In addition to results in the present scheme, $\rho_f^{\rm (S)}$ is shown for comparison. 
Characteristic energy scales turn out to increase as $U$ decreases. 
We define the Kondo temperature $T_{\rm K}$ by the peak position of the resonance, and show rescaled spectra as a function of $\omega /T_{\rm K}$ in Fig.~\ref{fig:dos_n6}(b). 
The spectra for $U\geq 3|\epsilon_f|$ are 
almost identical to each other, while the spectrum for $U=2|\epsilon_f|$ is a little 
different from the others. 
The excellent scaling property 
of the Kondo resonance ensures an accuracy of the calculated intensity in the present scheme.
On the other hand, $\rho_f^{\rm (S)}$ clearly fails to demonstrate the scaling  property. 
The vertex correction in the Green function 
consequently turns out essential to reproduce the universal behavior of the Kondo systems.
Spectra of $n=4$ also exhibit the scaling property as shown in Fig.~\ref{fig:dos_n4}.

\begin{figure}[t]
	\begin{center}
	\includegraphics[width=8.5cm]{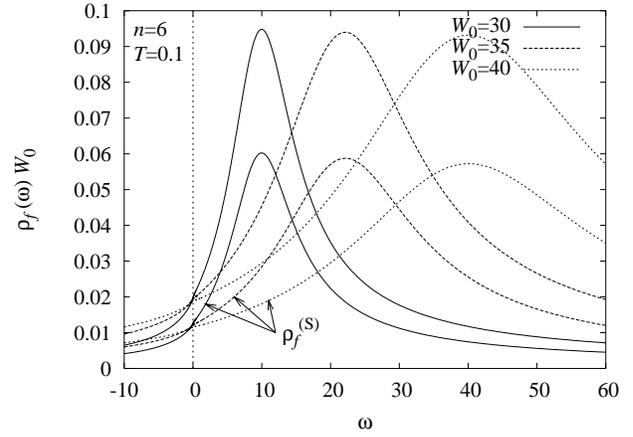}
	\end{center}
	\caption{Single-particle spectra near the Fermi level for several values of $W_0$ at $T=0.1$ with parameters $n=6$, $\epsilon_f =-1200$ and $U=5|\epsilon_f|=6000$.}
	\label{fig:dos_W}
\end{figure}

Figure~\ref{fig:dos_W} shows $W_0$ dependence of spectra with $n=6$ and $U=6000$. 
The ordinate is the density of states multiplied by $W_0$.
The peak hights in the scaled ordinate are nearly identical, which is also consistent
with the scaling property as expected.

\subsection{The Kondo temperature}
We employ the definition of $T_{\rm K}$ as the energy scale of the system at zero temperature.\cite{Hewson, Kuramoto-K} 
Namely the $4f$ magnetic susceptibility $\chi$ at $T=0$ determines $T_{\rm K}$ by
\begin{align}
 	\chi = Cn_f/T_{\rm K},
	\label{eq:chi}
\end{align}
where $C$ is the Curie constant. 
Assuming $n_f=1$ with frozen charge fluctuation,
the Fermi liquid theory gives
the $4f$ contribution to the specific heat coefficient $\gamma$ as
\begin{align}
	\gamma = \frac{\pi^2}{3T_{\rm K}} \frac{n-1}{n}.
	\label{eq:gamma}
\end{align}
In the Fermi liquid theory, the quasi-particle density of states $\rho_f^*(0)$ 
per spin-orbital channel at the Fermi level determines $\gamma$ as
$\gamma = n\pi^2 \rho_f^*(0)/3$.
On the other hand, the original
density $\rho_f(0)$ of $4f$ states is given exactly by the Friedel sum-rule as
\begin{align}
	\rho_f (0) = \frac{1}{\pi^2 W_0}\sin^2 \left( \frac{\pi n_f}{n} \right).
	\label{eq:rho0}
\end{align}
The wave function renormalization factor $a_f$ connects these densities of states by
\begin{equation}
\rho_f(0)= a_f\rho_f^*(0). 
\end{equation}

At zero temperature, an approximate analytic formula for $\rho_f(\omega)$ 
has been proposed \cite{Kuramoto-MH}, which interpolates the Fermi liquid result and the high-energy tail of the Kondo resonance.   
Between the Fermi level and the peak position of the Kondo resonance, the density of states is approximately given by \cite{Kuramoto-MH,Kuramoto-K}
\begin{equation}
\rho_f(\omega) = \frac{a_f}{\pi} \cdot
\frac{\tilde{\Delta}} {(\omega-\tilde{\epsilon}_f)^2+\tilde{\Delta}^2},
\end{equation}
where 
$\tilde{\epsilon}_f$ and $\tilde{\Delta}$ are the renormalized $4f^1$ level and its width, respectively.  
Requiring the same number of $4f$ electrons and $4f$ quasi-particles, we obtain the relation
\begin{equation}
\tilde{\Delta}^2/(\tilde{\epsilon}_f^2+\tilde{\Delta}^2)=
\sin^2 \left( \pi n_f/n \right).
\end{equation}
Hence the renormalization factor is given by $a_f = \tilde{\Delta}/(\pi W_0)$.
Furthermore comparison with eq.~(\ref{eq:gamma}) 
relates the peak position 
$\tilde{\epsilon}_f$  and $T_{\rm K}$ with $n_f=1$
by \cite{Hewson}
\begin{align}
\tilde{\epsilon}_f  = 
\frac{n^2 \sin(\pi /n) \cos(\pi /n)}{\pi (n-1)}T_{\rm K}.
\label{eq:peak}
\end{align}
Hence we can derive $T_{\rm K}$ from the peak position of $\rho_f(\omega)$.
The proportionality factor in eq.~(\ref{eq:peak})
is 0.99 for $n=6$ and 0.85 for $n=4$. 


We next compare the Kondo temperature $T_{\rm K}$ derived from $\rho_f(\omega)$ with analytic formula derived by the scaling theory. 
In the parameter regime such that $-\epsilon_f \gg D$ and $\epsilon_f +U \gg D$, 
a combination of the Schrieffer-Wolff transformation and the 
second-order scaling theory 
gives an analytic expression for the Kondo temperature of the form\cite{Hewson}
\begin{align}
	&T_{\rm K}^{(2)} = c D g^{1/n} \exp (-1/g), \nonumber \\
	&g = nW_0 [\ |\epsilon_f|^{-1} + (U +\epsilon_f)^{-1}],
	\label{eq:scaling}
\end{align}
where $c$ is a numerical constant of the order of unity, which cannot be determined by the scaling theory.
We fix $c$ as unity in this paper for simplicity.
If the parameters are such that $-\epsilon_f < D$ or $\epsilon_f +U < D$, 
conduction band states with excitation energies around $D$ 
do not participate in renormalization processes for the Kondo screening. 
Then the 
cut-off energies are given roughly by
$\max\{ -D,\epsilon_f\}$ for the lower and $\min\{ D,\epsilon_f +U\}$ for the upper.
In such a case,  estimate of $T_{\rm K}$ becomes more complicated.\cite{Haldane, Hewson} 
In order to avoid the complexity,
we solve integral equations under the Coqblin-Schrieffer limit, that is, $-\epsilon_f, U, W_0 \rightarrow \infty$ with 
$g$ fixed. 
Then eq.~(\ref{eq:scaling}) 
is valid for comparison with the present theory.
The limit 
is taken at the stage of integral equations by introducing 
auxiliary functions
such as $\tilde{J}_0 (z) = -W_0 R_0 (z)$ and $\Pi_0 (z) = \Sigma_0 (z)/W_0$ 
that remain finite under the limit operation 
\cite{Bickers}.

Table~\ref{tab:tk} shows $T_{\rm K}$ and $T_{\rm K}^{(2)}$, determined respectively by eqs.~(\ref{eq:peak}) and (\ref{eq:scaling}), for several values of parameters. 
%
\begin{table}[t]
	\begin{center}
	\caption{The Kondo temperature for various sets of parameters. $T_{\rm K}$ is a value determined from the peak position of the Kondo resonance by eq.~(\ref{eq:peak}), and $T_{\rm K}^{(2)}$ is that evaluated by eq.~(\ref{eq:scaling}).}
	\label{tab:tk}
	\begin{tabular}{ccccc}
	\multicolumn{5}{l}{(a) present theory ($n=6$,\ $W_0/|\epsilon_f|=0.025$)} \\
	\hline
	$\epsilon_f$  & $U /|\epsilon_f|$ & $T_{\rm K}$ & $T_{\rm K}^{(2)}$ & $T_{\rm K}/T_{\rm K}^{(2)}$  \\
	\hline
	$-\infty$ & 2 & 238 & 292 & 0.81 \\
	$-\infty$ & 3 & 77 & 92 & 0.84 \\
	$-\infty$ & 4 & 45 & 52 & 0.88 \\
	$-\infty$ & 5 & 33 & 37 & 0.90 \\
	$-\infty$ & $\infty$ & 10 & 9.3 & 1.09 \\
	\hline
	$-1200$ & 2 & 89 & -- & -- \\
	$-1200$ & 3 & 23 & -- & -- \\
	$-1200$ & 4 & 13 & -- & -- \\
	$-1200$ & 5 & 10 & -- & -- \\
	$-1200$ & $\infty$ & 5.5 & -- & -- \\
	\hline
	\end{tabular}
	\begin{tabular}{ccccc}
	\multicolumn{5}{l}{(b) present theory ($n=4$,\ $W_0/|\epsilon_f|=0.0375$)} \\
	\hline
	$\epsilon_f$  & $U /|\epsilon_f|$ & $T_{\rm K}$ & $T_{\rm K}^{(2)}$ & $T_{\rm K}/T_{\rm K}^{(2)}$  \\
	\hline
	$-\infty$ & 2 & 201 & 264 & 0.76 \\
	$-\infty$ & 3 & 67 & 80 & 0.83 \\
	$-\infty$ & 4 & 40 & 45 & 0.89 \\
	$-\infty$ & 5 & 30 & 32 & 0.94 \\
	$-\infty$ & $\infty$ & 9.9 & 7.9 & 1.25 \\
	\hline
	$-1200$ & 2 & 70 & -- & -- \\
	$-1200$ & 3 & 18 & -- & -- \\
	$-1200$ & 4 & 10 & -- & -- \\
	$-1200$ & 5 & 7.6 & -- & -- \\
	$-1200$ & $\infty$ & 4.2 & -- & -- \\
	\hline
	\end{tabular}
	\begin{tabular}{ccccc}
	\multicolumn{5}{l}{(c) NCA$f^2$ ($n=6$,\ $W_0/|\epsilon_f|=0.025$)} \\
	\hline
	$\epsilon_f$  & $U /|\epsilon_f|$ & $T_{\rm K}$ & $T_{\rm K}^{(2)}$ & $T_{\rm K}/T_{\rm K}^{(2)}$  \\
	\hline
	$-\infty$ & 2 & 7.9 & 292 & 0.027 \\
	$-\infty$ & 3 & 9.6 & 92 & 0.15 \\
	$-\infty$ & 4 & 9.8 & 52 & 0.19 \\
	$-\infty$ & 5 & 10 & 37 & 0.27 \\
	$-\infty$ & $\infty$ & 10 & 9.3 & 1.09 \\
	\hline
	\end{tabular}
	\end{center}
\end{table}
%
Table~\ref{tab:tk}(a) shows that the present theory gives 
a reasonable agreement 
with $T_{\rm K}^{(2)}$ in $n=6$ over a wide range of $U$. 
Deviations from $T_{\rm K}^{(2)}$ become larger in the case of small $U$.
We attribute the deviations to the approximation for the vertex function, which neglects energy dependence of 
the $4f^2$ resolvent in the renormalization processes.
In the case of $n=4$, on the other hand, 
variation of $T_{\rm K}/T_{\rm K}^{(2)}$ for different values of $U/|\epsilon_f|$
is larger than the case of  $n=6$.
This is mainly attributed to neglect of higher order terms 
in the $1/n$ expansion. 
Namely the present theory ($U<\infty$) neglects 
many contributions of order $1/n$ and beyond, while
the original NCA ($U=\infty$) keeps all $1/n$ order terms but neglects
some $1/n^2$ contributions.

We can see from Table~\ref{tab:tk}(a) and (b) that the $4f^0$ and $4f^2$ levels 
lying within the conduction band lead to quite lower values of $T_{\rm K}$ as compared to
the case of the Coqblin-Schrieffer limit. 
The reduction of $T_{\rm K}$ should be due to the smaller cut-off energies caused by the virtual charge excitations.

Table~\ref{tab:tk}(c) shows $T_{\rm K}$ evaluated with use of the NCA$f^2$. 
No enhancement of $T_{\rm K}$ by finite values of $U$ is obtained by this lowest order self-consistent approximation.
We thus conclude that 
inclusion of higher order terms of the vertex correction
is the essence of the present theory which
gives the proper energy scale of the Kondo effect.

\subsection{Sum-rules and exact relations}
We now examine whether the present theory satisfies the exact sum-rules.
Table~\ref{tab:sum} shows 
the integrated density of states
in the present theory determined by eqs.~(\ref{eq:4f_number}) and (\ref{eq:f2v_4f_number2}), 
together with the exact value given by eq.~(\ref{eq:dos_sum}).
\begin{table}[t]
	\begin{center}
	\caption{
Integrated density of states of $4f$ electrons in the present scheme evaluated by eqs.~(\ref{eq:4f_number}) and (\ref{eq:f2v_4f_number2}), and the exact value by eq.~(\ref{eq:dos_sum}), with $\epsilon_f =-1200$ and $T=0.1$. }
	\label{tab:sum}
	\begin{tabular}{ccc}
	\multicolumn{3}{l}{(a) $n=6$, $W_0 =30$} \\
	\hline
	$U /|\epsilon_f|$ &  (present theory) & (exact) \\
	\hline
	2 & 1.53 & 0.962 \\
	3 & 1.13 & 0.975 \\
	4 & 1.07 & 0.983 \\
	5 & 1.05 & 0.988 \\
	\hline
	\end{tabular}
	\begin{tabular}{ccc}
	\multicolumn{3}{l}{(b) $n=4$, $W_0 =45$} \\
	\hline
	$U /|\epsilon_f|$ &  (present theory) & (exact) \\
	\hline
	2 & 1.48 & 0.972 \\
	3 & 1.14 & 0.983 \\
	4 & 1.08 & 0.989 \\
	5 & 1.05 & 0.992 \\
	\hline
	\end{tabular}
	\end{center}
\end{table}
Note that the deviation from the exact value becomes larger as $U$ decreases,  while it is not affected by the value of $n$. 
We accordingly attribute the inaccuracy of 
integrated density of states
to the approximation of the vertex function, because 
the NCA$f^2$v scheme has a conserving property provided 
the integral equation (\ref{eq:vertex_eq}) is solved exactly. 
We further conclude that the large intensity comes from the $4f^2$ part at high 
energy regime above the Fermi level.
The conclusion is based on the following two reasons: 
first the particle number is not affected by the vertex correction, and secondly the Kondo resonance peak near the Fermi level shows the proper scaling behavior.
We have confirmed that the deviation from the sum rule becomes smaller at higher temperatures. 

Table~\ref{tab:rho0} shows $\rho_f(0)$ obtained in the present scheme at $T=0.1$ and the exact value at $T=0$ evaluated from $n_f$ by eq.~(\ref{eq:rho0}).
\begin{table}[t]
	\begin{center}
	\caption{The value $\rho_f(0) W_0$ obtained in the present theory at $T=0.1$, and the exact value at $T=0$ evaluated from $n_f$ by eq.~(\ref{eq:rho0}), with parameter $\epsilon_f =-1200$.}
	\label{tab:rho0}
	\begin{tabular}{cccc}
	\multicolumn{4}{l}{(a) $n=6$, $W_0 =30$} \\
	\hline
	$U /|\epsilon_f|$ &  (present theory) & (exact) & $n_f$ \\
	\hline
	2 & 0.0192 & 0.0220 & 0.926 \\
	3 & 0.0196 & 0.0228 & 0.944 \\
	4 & 0.0197 & 0.0228 & 0.944 \\
	5 & 0.0198 & 0.0219 & 0.942 \\
	$\infty$ & 0.0198 & 0.0219 & 0.925 \\
	\hline
	\end{tabular}
	\begin{tabular}{cccc}
	\multicolumn{4}{l}{(b) $n=4$, $W_0 =45$} \\
	\hline
	$U /|\epsilon_f|$ &  (present theory) & (exact) & $n_f$ \\
	\hline
	2 & 0.0388 & 0.0448 & 0.926 \\
	3 & 0.0391 & 0.0457 & 0.938 \\
	4 & 0.0392 & 0.0456 & 0.936 \\
	5 & 0.0392 & 0.0453 & 0.933 \\
	$\infty$ & 0.0391 & 0.0441 & 0.917 \\
	\hline
	\end{tabular}
	\end{center}
\end{table}
The deviations from the exact values are roughly 10 percent in 
$n=6$, and 15 percent in $n=4$.  The deviations are insensitive to the parameter $U$.  
Thus the present theory is not worse than the original NCA
in deriving $\rho_f(0)$.
It is known that a fictitious anomaly appears at the Fermi level due to the inaccuracy of the NCA at $T=0$\cite{nca3}. 
We can recognize tiny spikes at 
the Fermi level in Figs.~\ref{fig:dos_n6} and \ref{fig:dos_n4}. 
Note that the anomaly in $n=6$ is smaller than that in $n=4$. 
As in the original NCA,  the present theory does not 
give exact description of the Fermi liquid property.

\subsection{Thermodynamics}
We now present numerical results for thermodynamic quantities of $4f$ contribution, and then examine adequacy of the present theory for the Kondo effect.
The partition function $Z_f$ of $4f$ part is given in terms of the resolvents by eq.~(\ref{eq:part_func}). 
The entropy $S_f$ and specific heat $C_f$ of $4f$ contribution can be computed by numerical differentiations of $Z_f$ with resepct to temperature.
To obtain reasonable numerical accuracy, however, it requires computations at numerous values of temperature, especially for the second derivatives.
Following ref.~\citen{Otsuki}, we 
derive  temperature derivatives of resolvents by solving 
additional integral equations.  In this way 
thermodynamic quantities are computed without numerical differentiations. 

\begin{figure}
	\begin{center}
	\includegraphics[width=8.5cm]{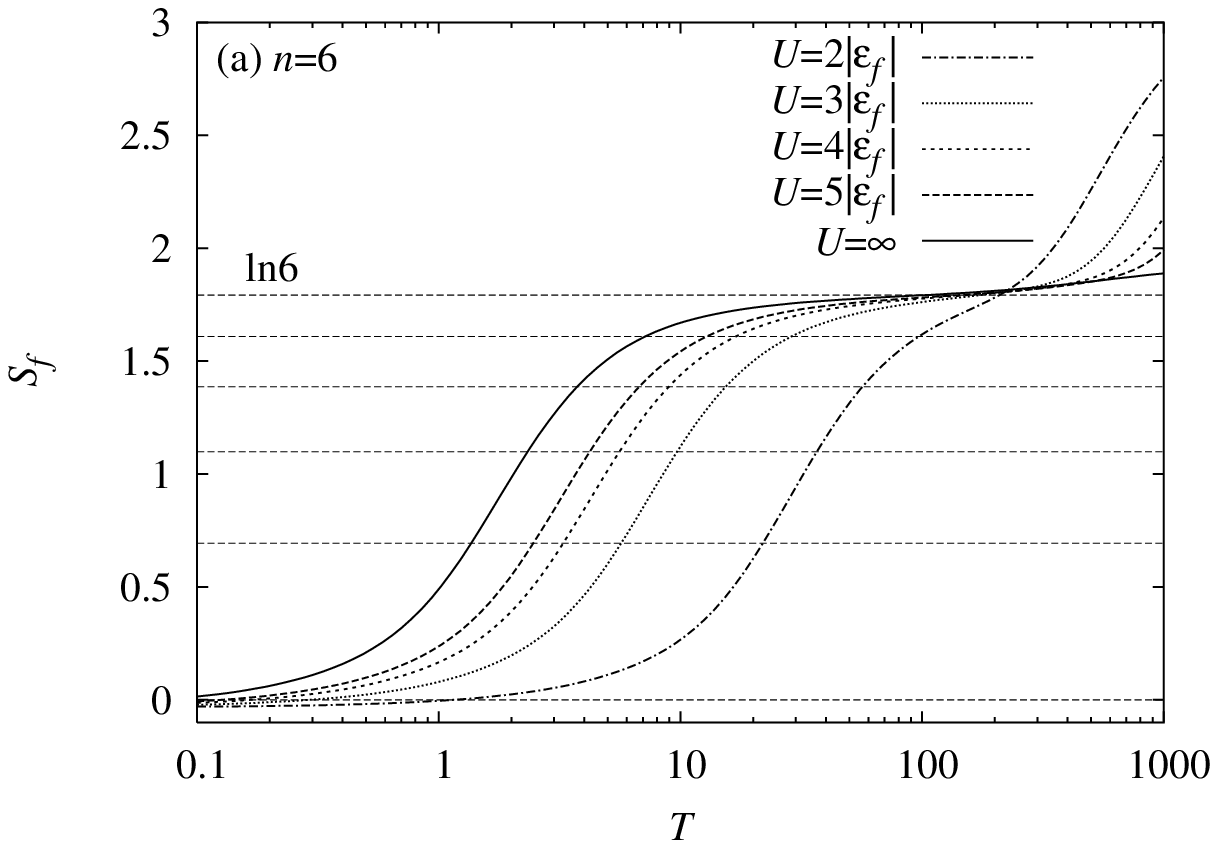}
	\includegraphics[width=8.5cm]{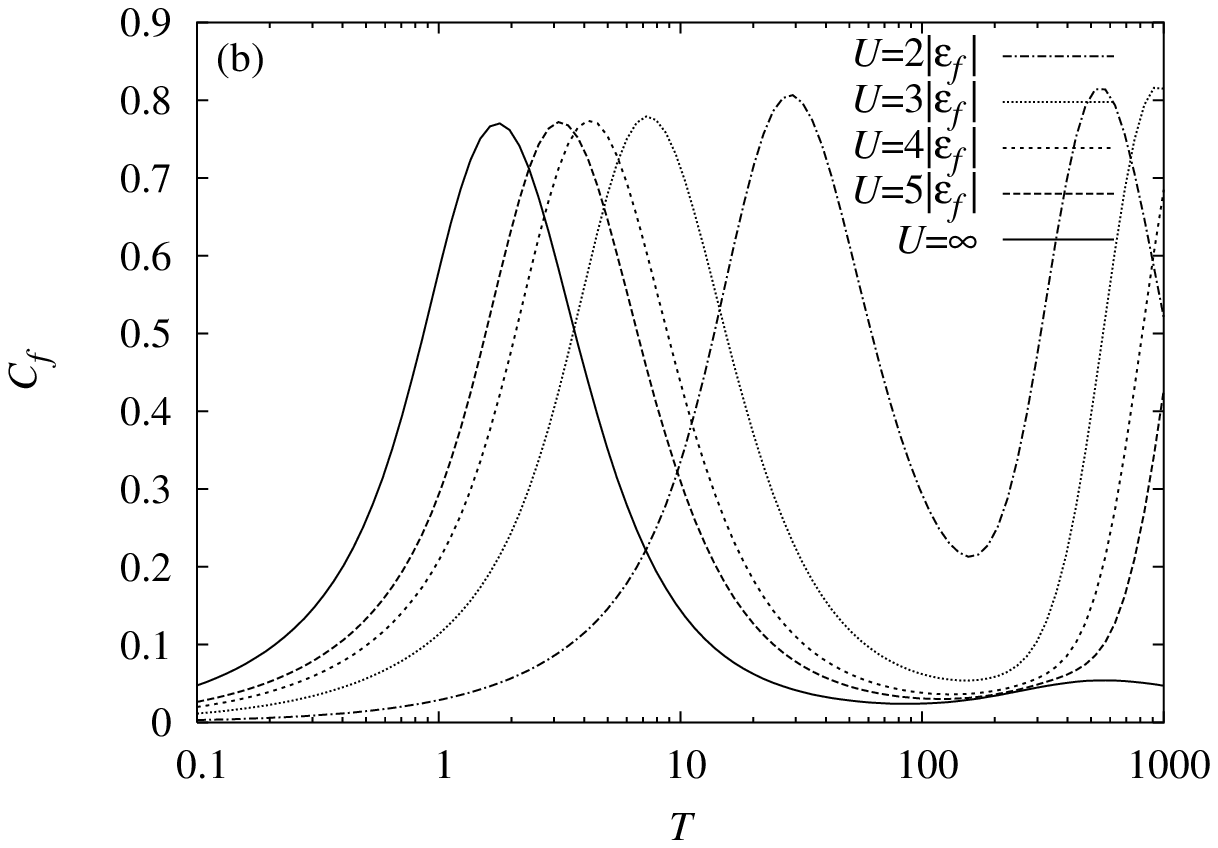}
	\includegraphics[width=8.5cm]{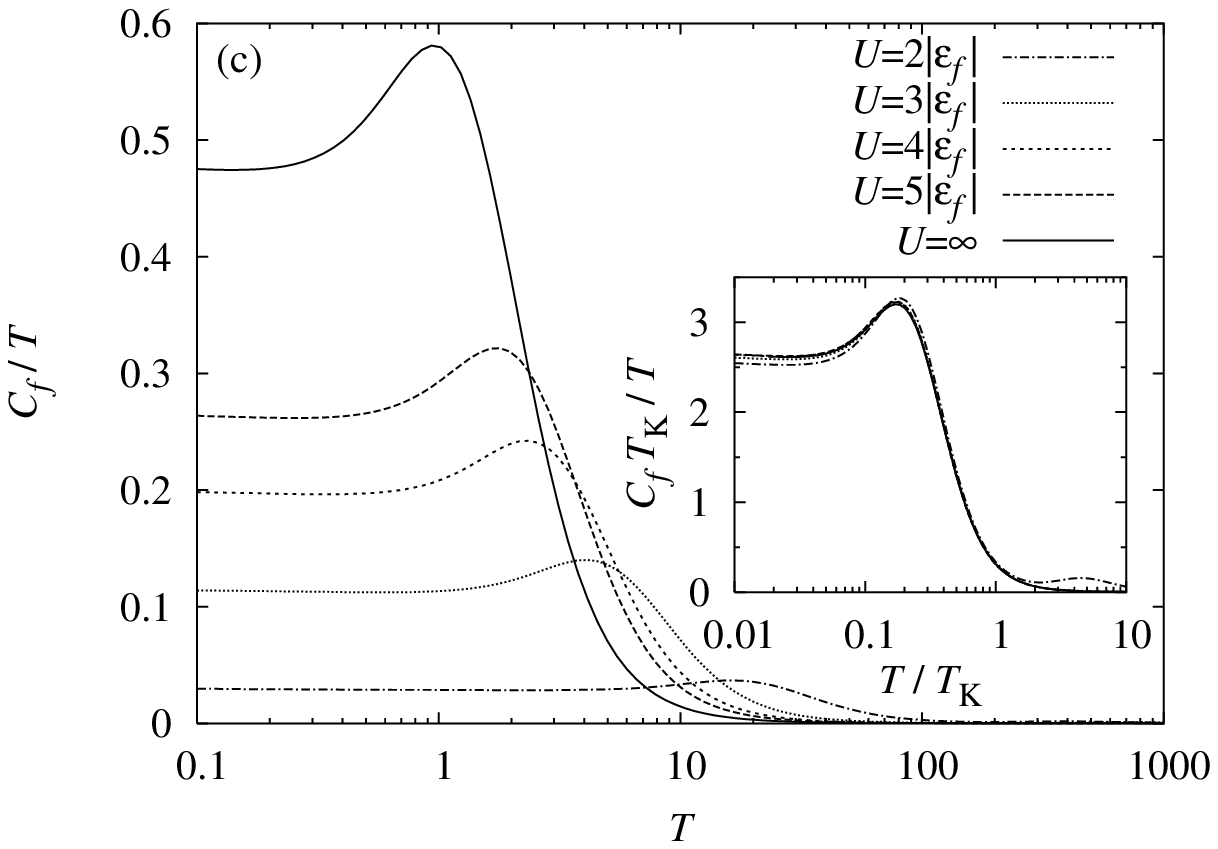}
	\end{center}
	\caption{Temperature dependence of the entropy $S_f$ and the specific heat $C_f$ of $4f$ contribution for several values of $U$ with $n=6$. Other parameters are same as Fig.~\ref{fig:dos_n6}. The values of $T_{\rm K}$ are listed in Table~\ref{tab:tk}(a).}
	\label{fig:thermo_n6}
\end{figure}

Figure~\ref{fig:thermo_n6}(a) shows the entropy as a function of temperature for several values of $U$ with $n=6$.
\begin{figure}
	\begin{center}
	\includegraphics[width=8.5cm]{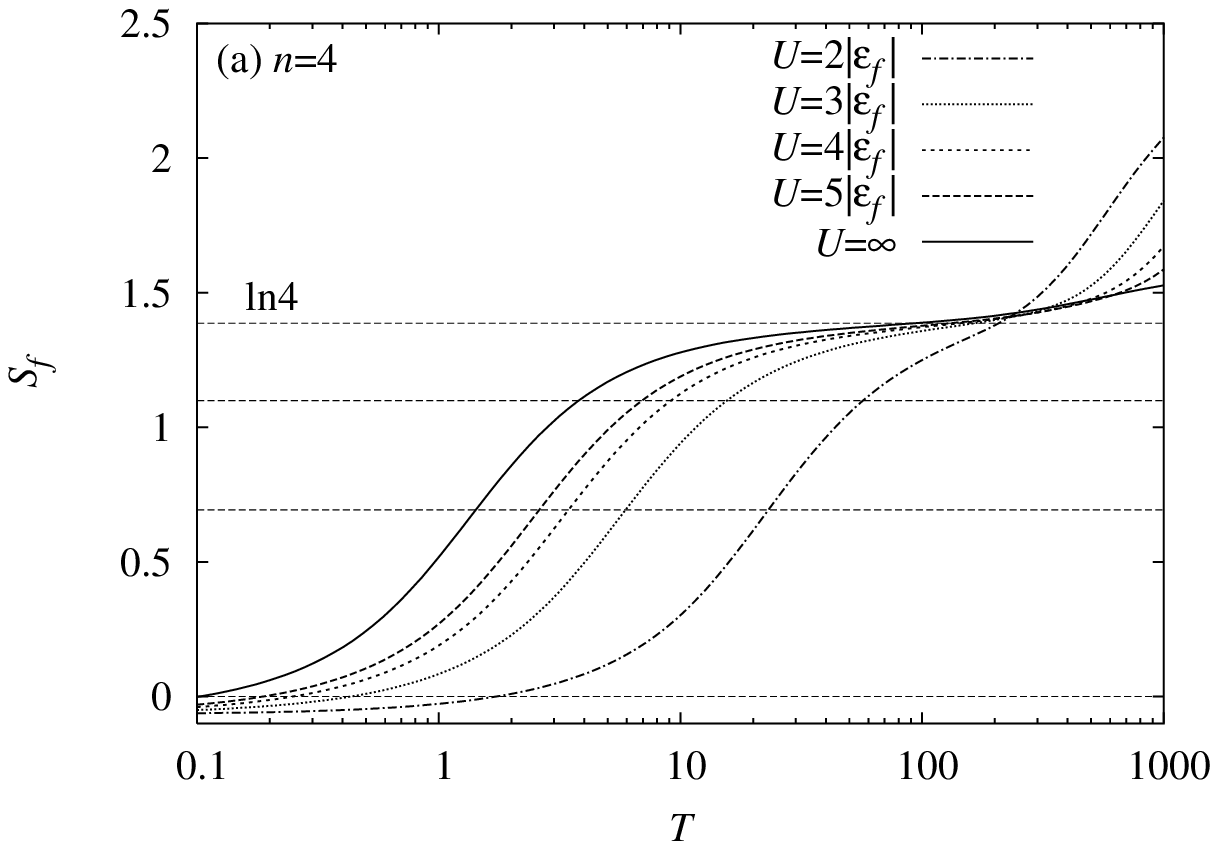}
	\includegraphics[width=8.5cm]{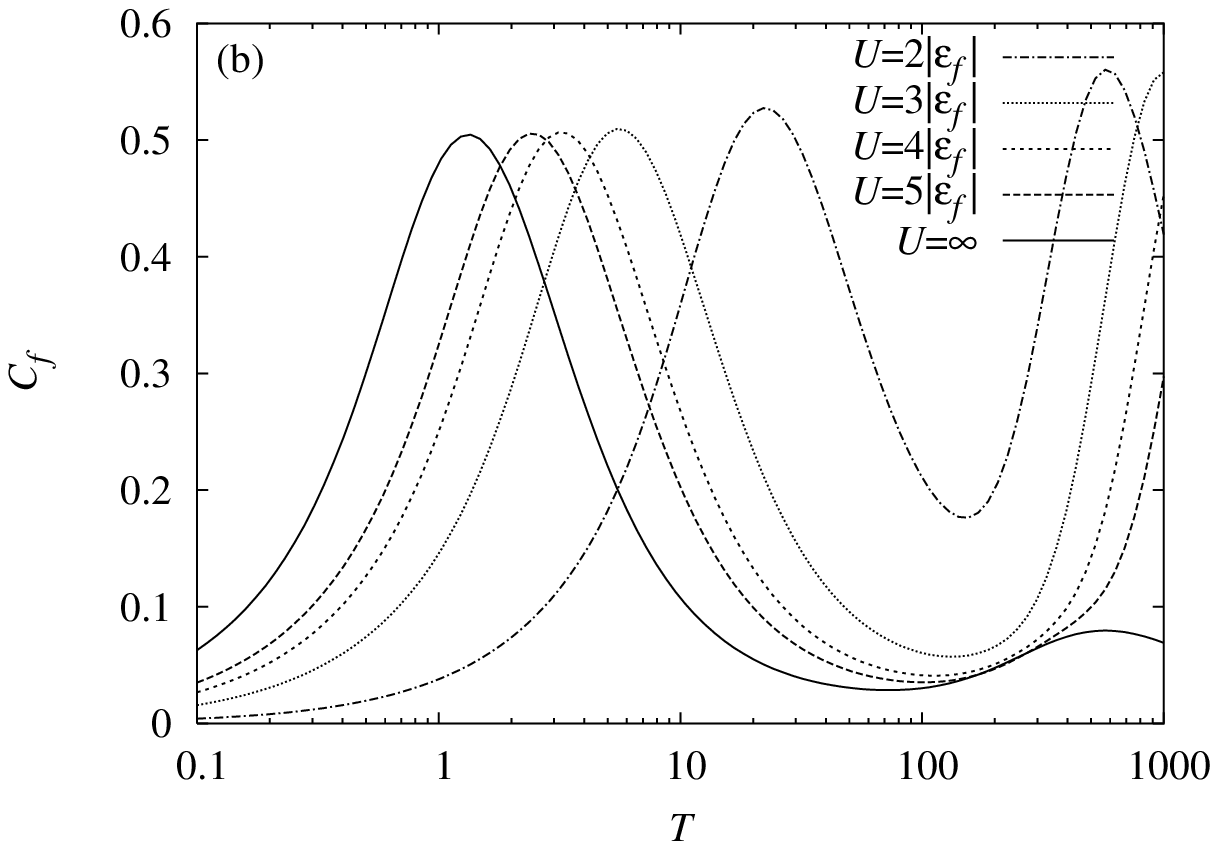}
	\includegraphics[width=8.5cm]{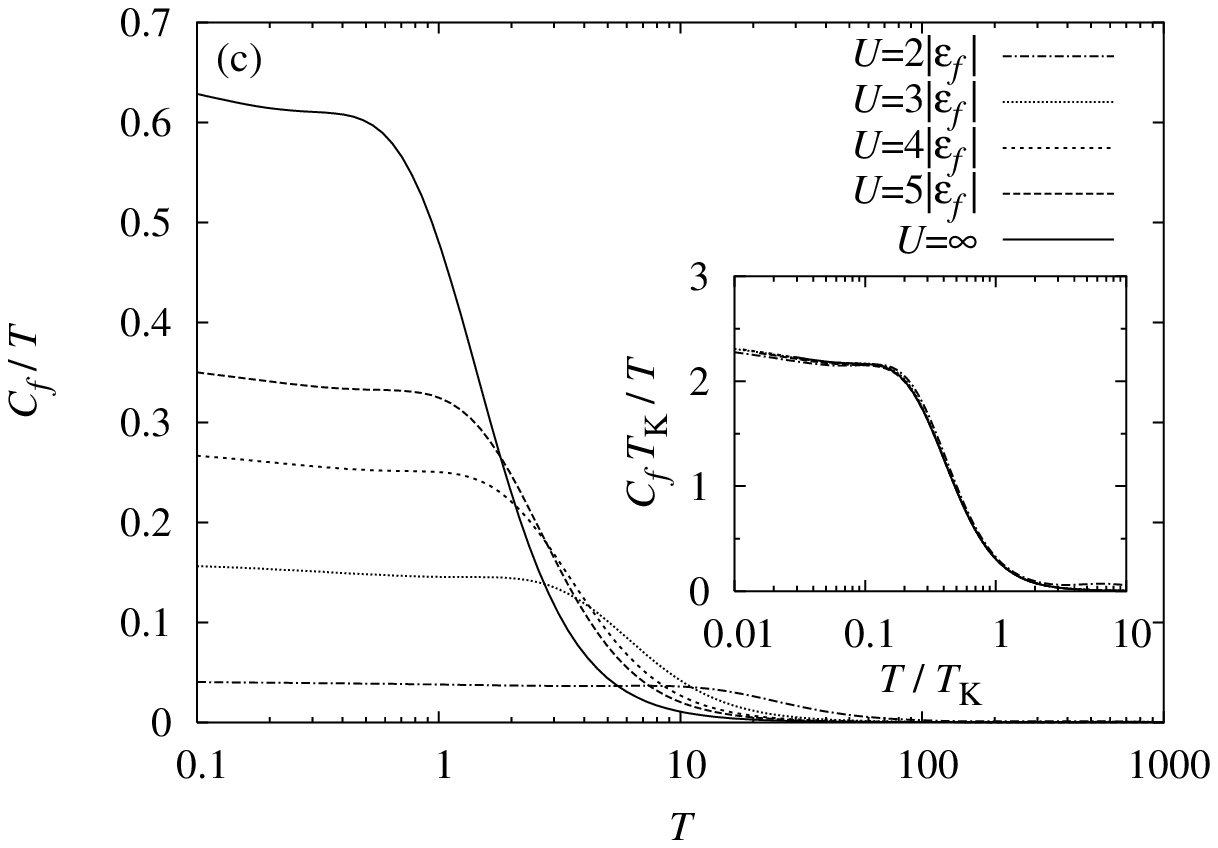}
	\end{center}
	\caption{Temperature dependence of the entropy $S_f$ and the specific heat $C_f$ of $4f$ contribution for several values of $U$ with $n=4$. Other parameters are same as Fig.~\ref{fig:dos_n4}. The values of $T_{\rm K}$ are listed in Table~\ref{tab:tk}(b).}
	\label{fig:thermo_n4}
\end{figure}
The entropy 
exhibits reduction from $\ln 6$ toward lower temperature due to the Kondo screening. 
The characteristic energy scale becomes large with decreasing $U$ as in the single-particle excitation spectra. 
At low temperatures, the entropy is slightly less than zero because of the inaccuracy of the NCA and the present theory. 
On the other hand, 
values larger than 
$\ln 6$ at high temperatures come from charge fluctuations involving $4f^0$ and $4f^2$ states.
Figures~\ref{fig:thermo_n6}(b) and (c) show temperature dependence of the specific heat $C_f$ and $C_f /T$, respectively. 
The $T$-linear behavior of $C_f$, 
namely $C_f /T$ being constant, 
is clearly seen at low temperatures. 
When temperature is scaled by $T_{\rm K}$ 
listed in Table~\ref{tab:tk}, $C_f /T$'s with different values of $U$ 
become almost identical to each other as shown in the inset of Fig.~\ref{fig:thermo_n6}(c).
Equation~(\ref{eq:gamma}) gives $\gamma T_{\rm K} \simeq 2.74$ for $n=6$, and is consistent with numerical results in the the inset of Fig~\ref{fig:thermo_n6}(c). 
Thus the present theory with $n=6$ demonstrates the scaling property of the Kondo effect in the specific heat as well as the single-particle spectrum.

On the other hand, thermodynamic quantities 
with $n=4$ show larger inaccuracy. 
The minus values of the entropy at low temperatures are seen more clearly than 
the $n=6$ case in Fig.~\ref{fig:thermo_n4}(a). 
Although the scaling behavior retains as shown in the inset of Fig.~\ref{fig:thermo_n4}(c), 
$C_f /T$ does not stay constant with decreasing temperature.
Thus in the case of $n=4$, the present theory (and the original NCA) has a limited accuracy to describe the specific heat at low temperatures.

\begin{figure}[t]
	\begin{center}
	\includegraphics[width=8.5cm]{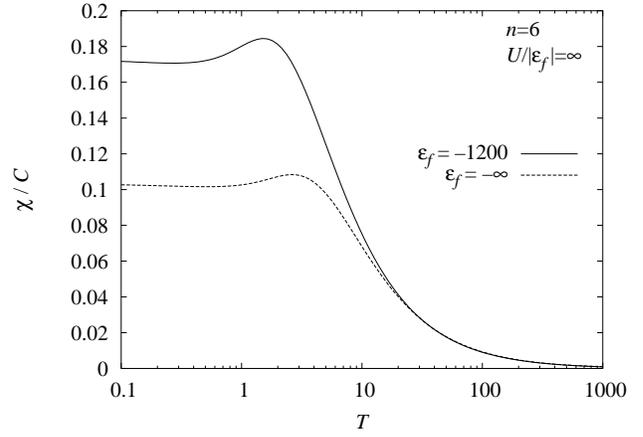}
	\end{center}
	\caption{Temperature dependence of the static magnetic susceptibility computed in the NCA with $n=6$ and $W_0 /|\epsilon_f |=0.025$. $C$ is the Curie constant.}
	\label{fig:suscep}
\end{figure}

We obtain the relation between $\chi$ and $\gamma$ by combining
eqs.~(\ref{eq:chi}) and (\ref{eq:gamma}) in the form
independent of $T_{\rm K}$ as follows:
\begin{align}
	\frac{\gamma C}{\chi} = \frac{\pi^2}{3} \frac{n-1}{n}.
	\label{eq:gamma_chi}
\end{align}
Derivation of the magnetic susceptibility for $U<\infty$ is much more complicated than the $U=\infty$ case because of the $4f^2$ contribution.  
Hence we check the accuracy in description of thermodynamics
by returning to the original NCA ($U=\infty$). 
Figure~\ref{fig:suscep} shows temperature dependence of the susceptibility computed in the NCA. 
From Figs.~\ref{fig:thermo_n6} and \ref{fig:suscep} we obtain $\gamma C/\chi =2.77$ at $T=0.1$, while eq.~(\ref{eq:gamma_chi}) gives 2.74. 
Thus the NCA ($U=\infty$) describes with a reasonable accuracy
the universal behavior in the specific heat and the static susceptibility. 

\section{Summary}
We have 
derived the single-particle spectrum within the conserving scheme, following the NCA$f^2$v approximation.
The theory incorporates infinite skeleton diagrams of leading contributions in the $1/n$ expansion. 
Equations have been simplified by approximations that are 
justifiable under the conditions of large-$n$ and large-$U$.
Resultant formulations have 
forms quite similar 
to those of the lowest order approximation: 
the equations for the Green function as well as the resolvents are obtained from the lowest order formulae
by only replacing $\Sigma_0 (z)$ by $\Sigma_0 (z) \Lambda (z)$, and $R_0 (z)$ by $R_0 (z) \Lambda (z)^2$.
The vertex function $\Lambda (z)$ has been given analytically.

Numerical calculations have shown that the scaling behavior of the Kondo resonance is well reproduced by the present theory. 
Thermodynamical quantities computed in the scheme are properly scaled by the characteristic energy as well. 
Furthermore, the Kondo temperature $T_{\rm K}$ has turned out to be consistent with analytic 
estimate for $n=6$ 
over a wide range of the Coulomb repulsion $U$. 
Although the present scheme breaks the sum-rule of the spectrum due to the simplification in actual calculations, 
the inaccuracy in intensity 
comes only from the $4f^2$ peak at high energy regime.
Thus the drawback does not affect the density of states in the vicinity of the Fermi level. 
In addition, the inaccuracy remains 
small as long as $U$ is so large that the relation $|\epsilon_f| < \epsilon_f +U$ holds. 
This is the case in  most of Ce compounds.

The present theory has the advantage that it hardly requires extra time for computations as compared to 
the lowest order self-consistent approximation. 
Consequently, the theory is easily applicable to the DMFT, and to models with 
complex $4f$ levels and realistic band structures.

\section*{Acknowledgment}
We acknowledge useful discussion with Prof. O. Sakai, and suggestion of Dr. H. Kusunose on the Kondo temperature.

\appendix
\section{Equations for the Defect Propagator}
Single-particle excitation has been represented in terms of two types of spectral functions 
$\eta_{\gamma}(\omega)$ and $\xi_{\gamma}(\omega)$ defined by eqs.~(\ref{eq:eta}) and (\ref{eq:xi}). 
Although $\xi_{\gamma}(\omega)$ is related to $\eta_{\gamma}(\omega)$ analytically, it is difficult to compute $\xi_{\gamma}(\omega)$ directly due to the Boltzmann factor. 
Following ref.~\citen{nca3}, we transform original equations for the resolvent into 
equivalent but more convenient 
forms, and calculate $\xi_{\gamma}(\omega)$ by numerical iterations.

In order to make expressions simple, we introduce an operator $\mathcal{P}$ defined by $\mathcal{P}R_{\gamma} (\omega) = -Z_f^{-1} \text{e}^{-\beta \omega} \pi^{-1} \text{Im} R_{\gamma}(\omega +{\rm i}\delta)$. 
With use of the operator $\mathcal{P}$, we define $\sigma_{\gamma}(\omega) = \mathcal{P} \Sigma_{\gamma}(\omega)$ and $\xi_{\Lambda}(\omega) = \mathcal{P}\Lambda(\omega)$.
Operating $\mathcal{P}$ on eqs.~(\ref{eq:resolv}) and (\ref{eq:vertex}), we obtain
\begin{align}
	\xi_{\gamma} (\omega) &= 
	 |R_{\gamma}(\omega +{\rm i}\delta)|^2 \sigma_{\gamma} (\omega), 
	\label{eq:xi_begin} \\
	\xi_{\Lambda} (\omega) &= 
	 (n-1) \tilde{R}_2
	|\Lambda (\omega+{\rm i}\delta)|^2 \tilde{\sigma}_0^{\rm (NCA)} (\omega),
\end{align}
where $\tilde{\sigma}_0^{\rm (NCA)}(\omega)$ is defined by $\tilde{\sigma}_0^{\rm (NCA)}(\omega) = \mathcal{P}\tilde{\Sigma}_0^{\rm (NCA)}(\omega)$ and is calculated by
\begin{align}
	\tilde{\sigma}_0^{\rm (NCA)} (\omega) = \int {\rm d}\epsilon W(\epsilon) f(-\epsilon) \xi_1 (\omega +\epsilon),
\end{align}
which follows from an operation of $\mathcal{P}$ on eq.~(\ref{eq:self_nca}).
Similarly, operations of $\mathcal{P}$ on eqs.~(\ref{eq:self0}), (\ref{eq:self1}) and (\ref{eq:self2}) yield 
equations for $\sigma_{\gamma}(\omega)$ as follows:
\begin{align}
	\sigma_0 (\omega) &=
	n\xi_{\Lambda} (\omega) \tilde{\Sigma}_0^{\rm (NCA)}(\omega +{\rm i}\delta) \nonumber \\
	 & \quad + n\Lambda (\omega +{\rm i}\delta)^* \tilde{\sigma}_0^{\rm (NCA)}(\omega), \\
	\sigma_1 (\omega) &=
	 \int {\rm d}\epsilon W(\epsilon) [
	 f(\epsilon) \tilde{\xi}_0 (\omega-\epsilon) \nonumber \\
	 & \quad + (n-1) f(-\epsilon)\xi_2 (\omega+\epsilon)], \\
	\sigma_2 (\omega) &=
	 2\int {\rm d}\epsilon W(\epsilon) f(\epsilon) \xi_1 (\omega-\epsilon),
\end{align}
where $\tilde{\xi}_0 (\omega)$ is defined by $\tilde{\xi}_0 (\omega)=\mathcal{P}\tilde{R}_0(\omega)$.
Equation~(\ref{eq:R0_tilde}) leads to an explicit expression
\begin{align}
	\tilde{\xi}_0 (\omega) &= \xi_0 (\omega) {\rm Re}[\Lambda(\omega +{\rm i}\delta)^2 ] \nonumber \\
	 &+ 2 {\rm Re}[R_0 (\omega +{\rm i}\delta)] {\rm Re}[\Lambda (\omega +{\rm i}\delta)] \xi_{\Lambda} (\omega).
	\label{eq:xi_end}
\end{align}
These linear equations, (\ref{eq:xi_begin})--(\ref{eq:xi_end}), do not determine norms of $\xi_{\gamma}(\omega)$, $\sigma_{\gamma}(\omega)$.
The norms can be determined by the following sum-rule:
\begin{align}
	\int \text{d}\omega \sum_{\gamma} \xi_{\gamma}(\omega) = 1.
	\label{eq:sum_xi}
\end{align}
This identity follows from eq.~(\ref{eq:part_func}). 
Normalizing by eq.~(\ref{eq:sum_xi}) is equivalent to determination of the partition function, and thus thermodynamics is obtained by a combination of $\eta_{\gamma}(\omega)$ and $\xi_{\gamma}(\omega)$\cite{Otsuki}.


\begin{thebibliography}{99}
\bibitem{Georges} A. Georges, G. Kotliar, W. Krauth and M. J. Rozenberg: Rev. Mod. Phys. \textbf{68} (1996) 13. 
\bibitem{LDA_DMFT1} I. A. Nekrasov, . Held, N. Bl\"umer, A. I. Poteryaev, V. I. Anisimov and D. Vollhardt: Eur. Phys. J. B \textbf{18} (2000) 55.
\bibitem{LDA_DMFT2} M. B. Z\"olfl, I. A. Nekrasov, Th. Pruschke, V. I. Anisimov and J. Keller: Phys. Rev. Lett. \textbf{87} (2001) 276403.
\bibitem{Sakai2} O. Sakai, Y. Shimizu and Y. Kaneta: J. Phys. Soc. Jpn. {\bf 74} (2005) 2517.
\bibitem{nca1} Y. Kuramoto: Z. Phys. B \textbf{53} (1983) 37; H. Kojima, Y. Kuramoto and M. Tachiki: Z. Phys. B \textbf{53} (1984) 293.
\bibitem{Grewe} N. Grewe: Z. Phys. B \textbf{53} (1983) 271.
\bibitem{Keiter-Morandi} H. Keiter and G. Morandi: Phys. Rep. \textbf{109} (1984) 227.
\bibitem{Bickers} N. E. Bickers: Rev. Mod. Phys. \textbf{59} (1987) 845.
\bibitem{Coleman} P. Coleman: Phys. Rev. B {\bf 29} (1984) 29.
\bibitem{Wilson} K. G. Wilson: Rev. Mod. Phys. \textbf{47} (1975) 773.
\bibitem{Hirsch-Fye} J. E. Hirsch and R. M. Fye: Phys. Rev. Lett. \textbf{56} (1986) 2521.
\bibitem{xnca1} Y. Kuramoto: in {\it Theory of Heavy Fermions and Valence Fluctuations}, ed. T. Kasuya and T. Saso (Springer-Verlag, Berlin, 1985) p. 152.
\bibitem{xnca2} C.-I. Kim, Y. Kuramoto and T. Kasuya: J. Phys. Soc. Jpn. \textbf{59} (1990) 2414.
\bibitem{Pruschke_Hubbard} Th. Pruschke, D. L. Cox and M. Jarrell: Phys. Rev. B {\bf 47} (1993) 3553.
\bibitem{Sakai1} O. Sakai, M. Motizuki and T. Kasuya: in {\it Core-Level Spectroscopy in Condensed Systems Theory}, ed. J. Kanamori and A. Kotani (Springer-Verlag, Berlin, 1988) p. 45.
\bibitem{Pruschke} Th. Pruschke and N. Grewe: Z. Phys. B {\bf 74} (1989) 439.
\bibitem{Saso} T. Saso: Prog. Theor. Phys. Suppl. {\bf 108} (1992) 89.
\bibitem{Kroha} K. Haule, S. Kirchner, J. Kroha and P. W\"olfle: Phys. Rev. B {\bf 64} (2001) 155111; 
 J. Kroha and P. W\"olfle: J. Phys. Soc. Jpn. {\bf 74} (2005) 16.

\bibitem{Kuramoto-MH} 
Y. Kuramoto and E. M\"{u}ller-Hartmann: J. Magn. Magn. Mater. {\bf 52} (1985) 122.
\bibitem{Kuramoto-K} 
Y. Kuramoto and Y. Kitaoka: {\it Dynamics of Heavy Electrons} (Oxford University Press, 2000).
\bibitem{Hewson} 
A. C. Hewson: {\it The Kondo Problem to Heavy Fermions} (Cambridge University Press, 1993).
\bibitem{Haldane} F. D. M. Haldane: J. Phys. C \textbf{11} (1978) 5015.
\bibitem{nca3} Y. Kuramoto and H. Kojima: Z. Phys. B \textbf{57} (1984) 95.
\bibitem{Otsuki} J. Otsuki, H. Kusunose and Y. Kuramoto: cond-mat/0510424.

\end{thebibliography}
\end{document}